\documentclass[pra,aps,english,twocolumn,superscriptaddress,showpacs,showkeys]{revtex4-1}
\usepackage{babel,calc,amsmath,amsthm,amssymb,graphicx,subfigure,xcolor,comment}
\usepackage{mathdots}
\usepackage[T1]{fontenc}
\usepackage[unicode=true]{hyperref}
\hypersetup{
	colorlinks=true,       		
	linkcolor=blue,          	
	citecolor=red,            
	urlcolor=magenta,           	
}

\begin{document}

	\title{Deriving Einstein-Podolsky-Rosen steering inequalities from the few-body Abner Shimony inequalities}
	\author{Jie Zhou}
	\affiliation{Theoretical Physics Division, Chern Institute of Mathematics, Nankai University, Tianjin 300071, People's Republic of China}
	
	\author{Hui-Xian Meng}
	\affiliation{Theoretical Physics Division, Chern Institute of Mathematics, Nankai University, Tianjin 300071, People's Republic of China}
	
	\author{Shu-Han Jiang}
	\affiliation{Theoretical Physics Division, Chern Institute of Mathematics, Nankai University, Tianjin 300071, People's Republic of China}
	\affiliation{School of Physics, Nankai University, Tianjin 300071, People's Republic of China}
	
	\author{Zhen-Peng Xu}
	\affiliation{Theoretical Physics Division, Chern Institute of Mathematics, Nankai University, Tianjin 300071, People's Republic of China}
	\affiliation{Departamento de F\'{\i}sica Aplicada II, Universidad de  Sevilla, E-41012 Sevilla, Spain}

	\author{Changliang Ren}
	\affiliation{Center for Nanofabrication and System Integration, Chongqing Institute of Green and Intelligent Technology, Chinese Academy of Sciences, People's Republic of China}
	\affiliation{Chongqing Key Laboratory of Multi-Scale Manufacturing Technology, Chongqing Institute of Green and Intelligent Technology, Chinese Academy of Sciences, People's Republic of China}
	
	\author{Hong-Yi Su}
	\email{hysu@mail.nankai.edu.cn}
	\affiliation{Theoretical Physics Division, Chern Institute of Mathematics, Nankai University, Tianjin 300071, People's Republic of China}
	
	\author{Jing-Ling Chen}
	\email{chenjl@nankai.edu.cn}
	\affiliation{Theoretical Physics Division, Chern Institute of Mathematics, Nankai University,
		Tianjin 300071, People's Republic of China}
	\affiliation{Centre for Quantum Technologies, National University of Singapore,
		3 Science Drive 2, Singapore 117543}
	
	\date{\today}
	\begin{abstract}
		For the Abner Shimony (AS)  inequalities, the simplest unified forms of directions attaining the maximum quantum violation  are investigated. Based on these directions,  a family of Einstein-Podolsky-Rosen (EPR) steering inequalities is derived from the AS inequalities  in a systematic manner. For these inequalities, the local hidden state (LHS) bounds are strictly less than the local hidden variable (LHV) bounds. This  means that the EPR steering is
		a form of quantum nonlocality strictly weaker than Bell-nonlocality.
	\end{abstract}
	
	
	
	\maketitle
	
	\section{Introduction}

	Quantum
	entanglement  distinguishes quantum theory from classical theory. With entangled states, the first authors to identify an interesting nonlocal effect
	were Einstein, Podolsky,
	and Rosen (EPR) in 1935 \cite{EPR}.  Subsequently, the concept of \textit{steering} was first introduced in 1935 by Schr$\ddot{\rm o}$dinger \cite{Schrodinger} as a reply to the Einstein-Podolsky-Rosen \cite{EPR} paradox. EPR steering reflects  a ``spooky action'' feature that manipulating one object seemingly affects another instantaneously, even
	it is far away. Wiseman pointed out in \cite{Wiseman} that steerability is stronger than entanglement but it is weaker than Bell nonlocality. Different from entanglement and Bell nonlocality,
	quantum steering is asymmetric between the two
	parties. In details, It may happen  that Alice can steer Bob but Bob
	can never steer Alice. This distinguished feature would be useful for some one-way quantum information tasks, such as quantum
	cryptography \cite{one-way-steer}.
	
	In 1964, Bell proposed an famous inequality for local hidden variable (LHV) models  \cite{Bell} to refute EPR paradox. Bell inequalities revealed that  quantum mechanics is incompatible with local realism. The more Bell inequalities
	we know, the more we know about the boundaries between
	Einstein's local realism and the genuinely nonclassical areas
	of quantum physics, which are potentially useful in quantum
	information applications. For instance, Bell inequalities have
	gained a utilitarian power in different quantum information
	tasks, such as quantum key distribution \cite{Ekert}, communication
	complexity \cite{Brukner}, and recently random number generation \cite{Pironio}.

	Similar to Bell inequalities, steering inequalities \cite{wisemanNature} have been proposed to reveal the EPR steerability of quantum states. It is in principle easier to experimentally observe the violation than Bell inequalities,
	because  one has no concerns about closing the notorious locality loophole as in a Bell test~\cite{Brunner}. Therefore, there is an important research significance in theory \cite{Chen,Skrzypczyk,Bowles} and experiment \cite{wisemanNature,Sun}. Based on the research approaches in the field of Bell's nonlocality, we  have   constructed chained steering inequalities from the chained Bell inequalities \cite{Meng}. Since the more steering inequalities we know, the more steerable states can be detected. In this paper, we focus on deriving EPR steering inequalities from the Abner Shimony (AS) inequalities  introduced in Ref. \cite{Gisin}.
	
	Without loss of generality, we will take an Alice-to-Bob steering scenario where
	correlations between classical variables declared by Alice but quantum expectation values found by Bob, in this sense~\cite{wisemanNature}. In fact, Bob's directions are taken  as those that can maximally violate the AS inequalities. Then a family of  steering inequalities are constructed. Finally,   a comparison between their local hidden state (LHS) bound and  their quantum violation is made in a systematical manner. Thus,  based on the comparison, we are able to compare the EPR steering  with the Bell nonlocality.

	The paper is organized as follows. In Sec.~\ref{AS}, we shall be briefly reviewing the AS inequalities and research the corresponding directions taken by Alice and Bob for every $N$, for which the maximum quantum values can be obtained. In Sec.~\ref{AS steering}, we will derive EPR-steering inequalities from the Abner Shimony inequalities, and compute the LHS bounds. By comparing the LHS bounds and the LHV bounds, we can find  that EPR-steering is a form of quantum nonlocality strictly weaker than Bell-nonlocality. Conclusions and discussion are put in the end of the paper.

	\section{The AS inequalities}\label{AS}
	
	The AS inequalities are a new family of tight bipartite Bell inequalities for any even number $N$ of inputs and binary outcomes, generalizing a tight  inequality
	introduced in Ref. \cite{Avis}. For binary inputs, the AS inequality is nothing but the Clauser-Horne-Shimony-Holt (CHSH) inequality \cite{chsh} (see Eq. (\ref{eqI2})). For ternary inputs, there is no new correlation inequality \cite{Collins,Avis}.  For any even number $N$ of settings with two possible outcomes, the AS Bell inequalities  can be written as~\cite{Gisin}
	
	\begin{align}\label{eq2140}
		I_N\overset{{\rm LHV}}{\leq}\mathcal{C}^N_{\rm LHV},
	\end{align}
	here
	\begin{align}\label{IN}
		I_N=\sum_{i,j}^{N}AS_N[i,j] \langle A_i B_j \rangle,
	\end{align}
	is the Bell expression, where  $AS_N[i,j]$ is the $i$th row and $j$th  column element of the following matrix $AS_N,$  $\langle A_iB_j\rangle$ denotes the mean value of the product of the
	outcomes of $A_i$ (measured on Alice's particle) and $B_j$ (measured on Bob's particle), and \begin{align}\label{Clhv}
		\mathcal{C}^N_{\rm LHV}=\frac{N}{2}\left(\frac{N}{2}+1\right),
	\end{align}
	is the classical bound for LHV models. The bound is obtained straightforwardly by the definition of the LHV models, i.e., by numerating all possible values $A_i,B_j=\pm1$ in $I_N$.

	\begin{widetext}
		
		\begin{align}\label{eqcij}
			AS_N=\left(
			\begin{array}{c||ccccccccccc}
				& A_1 &A_2 &A_3 & \cdots &A_n &A_{n+1} &A_{n+2} & \cdots &A_{2n-2} &A_{2n-1} & A_{2n} \\
				\hline
				\hline
				B_1     & 1   & 1   & 1   & \cdots & 1   & 1       & 1       & \cdots & 1        & 1        & 1 \\
				B_2     & 1   & 1   & 1   & \cdots & 1   & 1       & 1       & \cdots & 1        & 1        & -1 \\
				B_3     & 1   & 1   & 1   & \cdots & 1   & 1       & 1       & \cdots & 1        & -2       & 0 \\
				\vdots  & \vdots& \vdots &\vdots& \cdots & \vdots& \vdots &\vdots& \vdots  & \iddots        &\iddots       & \vdots \\
				B_n     & 1   & 1   & 1   & \cdots & 1   & 1       & -(n-1)  & \cdots & 0        & 0        &0 \\
				B_{n+1} & 1   & 1   & 1   & \cdots & 1   & -n      & 0       & \cdots & 0        & 0        &0 \\
				B_{n+2} & 1   & 1   & 1   & \cdots & -(n-1)& 0       & 0  & \cdots & 0        & 0        &0 \\
				\vdots  & \vdots& \vdots &\vdots& \cdots & \vdots& \vdots &\vdots& \vdots     & \vdots &\vdots& \vdots \\
				B_{2n-2}& 1   & 1   & 1    & \iddots & 0        & 0        &0      & \cdots & 0 & 0 & 0 \\
				B_{2n-1}& 1   & 1   & -2    & \iddots & 0        & 0        &0       & \cdots      & 0 & 0 & 0 \\
				B_{2n}  & 1   & -1   & 0    & \cdots & 0        & 0        &0       &\cdots      & 0 & 0 & 0\\
			\end{array}
			\right).
		\end{align}
	\end{widetext}

	For any  $N\geq2$ and $i=1,2,\cdots,N,$ if Alice and Bob  choose  the following directions
	\begin{align}\label{eqai}
		a_i=\left(\sin\theta_{a_i}\cos\phi_{a_i}, \sin\theta_{a_i}\sin\phi_{a_i},
		\cos\theta_{a_i}\right)
	\end{align}
	and
	\begin{align}\label{eqbi}
		b_i=\left(\sin\theta_{b_i}\cos\phi_{b_i}, \sin\theta_{b_i}\sin\phi_{b_i},
		\cos\theta_{b_i}\right)\end{align}
	respectively,
	then the observables that Alice and Bob choose are
	\begin{align}\label{eqAi}
		A_i&= \vec{\sigma}\cdot a_i\nonumber\\
		&=\sin\theta_{a_i}\cos\phi_{a_i}\sigma_x+\sin\theta_{a_i}\sin\phi_{a_i}\sigma_y+
		\cos\theta_{a_i}\sigma_z
	\end{align}
	and
	\begin{align}\label{eqBi}
		B_i&= \vec{\sigma}\cdot b_i\nonumber\\
		&=\sin\theta_{b_i}\cos\phi_{b_i}\sigma_x+\sin\theta_{b_i}\sin\phi_{b_i}\sigma_y+
		\cos\theta_{b_i}\sigma_z
	\end{align}
	respectively,
	where $\sigma_x,\sigma_y,\sigma_z$ are the Pauli matrices. Thus, the outcomes of $A_i$ and $B_j$ are either $1$ or $-1$.
	
	Assume that the initial quantum state of the compound system $\mathbb{C}^2\otimes\mathbb{C}^2$ is in the Spin singlet state
	\begin{equation}\label{eq0110}
		|\psi\rangle=\frac{1}{\sqrt{2}}(|01\rangle -|10\rangle),
	\end{equation}
	here $|0\rangle$ and $|1\rangle$ are eigenstates of $\sigma_z$ with eigenvalues 1 and $-1$, respectively.
	Then
	\begin{align}
		&~~~~\langle
		A_iB_j\rangle ={\rm tr}(A_iB_j|\psi\rangle\langle\psi|).
	\end{align}
	By Ref. \cite{Avis},
	the maximum quantum value of $I_N$ is
	\begin{align}\label{Inmax}
		I_{N\ {\rm{max}}}^{\rm QM}=\dfrac{(N+1)\sqrt{N(N+2)}}{3}.
	\end{align}
	Since the directions that  Alice and Bob choose to attain the maximum quantum value are not unique, we assume that
	Alice and Bob  choose the  measurement directions with the following forms:
	\begin{align}\label{a_i1}
		a_i=\left\{
		\begin{array}{lllll}
			(Y,\sqrt{1-Y^2}\sin\phi_0,\sqrt{1-Y^2}\cos\phi_0)~~~~~~i=1, \\
			(-Y,\sqrt{1-Y^2}\sin\phi_0,\sqrt{1-Y^2}\cos\phi_0)~~~~i=2\\
			(0,\sin\theta_{i-2},\cos\phi_{i-2})~~~~~~~~~~~~~~~~~~~~3 < i < N - 1\\
			(1,0,0)~~~~~~~~~~~~~~~~~~~~~~~~~~~~~~~~~~~~~~~~~~i=N,
		\end{array}
		\right.
	\end{align}
	\begin{align}\label{b_i1}
		b_i=\left\{
		\begin{array}{lllll}
			(Y,\sqrt{1-Y^2}\sin\theta_0,\sqrt{1-Y^2}\cos\theta_0)~~~~~~i=1, \\
			(-Y,\sqrt{1-Y^2}\sin\theta_0,\sqrt{1-Y^2}\cos\theta_0)~~~~i=2\\
			(0,\sin\theta_{i-2},\cos\theta_{i-2})~~~~~~~~~~~~~~~~~~~~3 < i < N - 1\\
			(1,0,0)~~~~~~~~~~~~~~~~~~~~~~~~~~~~~~~~~~~~~~~~~~i=N,
		\end{array}
		\right.
	\end{align}
	here \[Y=\frac{1}{\sqrt{\frac{N}{2}(\frac{N}{2}+1)}}.\]
	To the best of our knowledge, directions (\ref{a_i1}) and (\ref{b_i1}) are the simplest unified forms for $N=4,6,8,10$ to attain the maximum quantum value.
	
	In details, we list as follows some case studies:
	\begin{itemize}
		\item $N$=2: The coefficient matrix $AS_2$ is
		\begin{align}\label{eqI2}
			AS_2&=\left(
			\begin{array}{c||cc}
				& A_1 & A_2 \\
				\hline
				\hline
				B_1     & 1   & 1 \\
				B_2     & 1   & -1 \\
			\end{array}
			\right),
		\end{align}
		and so the AS inequality is nothing but the CHSH inequality, i.e.,
		\begin{equation*}
			\begin{aligned}
				I_2&=\sum_{i,j}^{2}AS_2[i,j] \langle A_i B_j \rangle\\
				&\equiv\langle A_1B_1\rangle+\langle A_1B_2\rangle+\langle A_2B_1\rangle-\langle A_2B_2\rangle\nonumber\\
				&\leq 2=\mathcal{C}_{\rm LHV}^2.
			\end{aligned}
		\end{equation*}

		If Alice and Bob fix their directions as
		\begin{align}\label{a_21}
			\begin{array}{lllll}
				a_1=(0,\sin\phi_0,\cos\phi_0),\\
				a_2=(0,\sin\phi_1,\cos\phi_1),
			\end{array}
		\end{align}
		and
		\begin{align}\label{b_41}
			\begin{array}{lllll}
				b_1=(0,\sin\theta_0,\cos\theta_0),\\
				b_2=(0,\sin\theta_1,\cos\theta_1),
			\end{array}
		\end{align}
		respectively,
		where \begin{equation}\label{b2}
			\theta_0=0,\theta_1=\pi,\phi_0=-\phi_1=-\dfrac{\pi}{2}-\arccos {\dfrac{1}{\sqrt{2}}},
		\end{equation}
		then the maximum quantum value  $I_{2\ {\rm{max}}}^{\rm QM}=2\sqrt{2}$ is obtained.

		\item $N$=4: The coefficient matrix $AS_4$ is
		\begin{align}\label{eqI4}
			AS_4&=\left(
			\begin{array}{c||cccc}
				& A_1 & A_2 & A_3 & A_4  \\
				\hline
				\hline
				B_1     & 1   & 1   & 1   & 1  \\
				B_2     & 1   & 1   & 1   & -1   \\
				B_3     & 1   & 1   & -2  & 0   \\
				B_4     & 1   & -1  & 0   & 0   \\
			\end{array}
			\right),
		\end{align}
		and so the Bell expression $I_4$ is
		\begin{equation}\label{I4}
			\begin{aligned}
				I_4&=\sum_{i,j}^{4}AS_4[i,j] \langle A_i B_j \rangle\\
				& \equiv \langle A_1B_1\rangle+\langle A_1B_2\rangle+\langle A_1B_3\rangle+\langle A_1B_4\rangle\nonumber\\
				&+\langle A_2B_1\rangle+\langle A_2B_2\rangle+\langle A_2B_3\rangle
				-\langle A_2B_4\rangle\nonumber\\
				&+\langle A_3B_1\rangle+\langle A_3B_2\rangle-2\langle A_3B_3\rangle\nonumber\\
				&+\langle A_4B_1\rangle-\langle A_4B_2\rangle\nonumber\\
				&\leq 6=\mathcal{C}_{\rm LHV}^4.
			\end{aligned}
		\end{equation}
		If Alice and Bob choose the measurement directions with forms as (\ref{a_i1}) and (\ref{b_i1}), with
		\[Y=\frac{1}{\sqrt{\frac{N}{2}(\frac{N}{2}+1)}}=\frac{1}{{\sqrt 6 }},\]
		and
		\begin{align}\label{b4}
			\begin{array}{l}
				\theta_1 = \frac{1}{2}\arccos \left[ {\frac{{ - 5}}{{3\sqrt 6 }}} \right],\theta_0 = \arccos \left[ {\frac{4}{{3\sqrt 5 }}} \right] - \theta_1,\\
				\phi_0 = \arccos \left[ {\frac{{ - 4}}{{3\sqrt 5 }}} \right] + \theta_1,\phi_1 = \arccos \left[ {\frac{5}{{3\sqrt 6 }}} \right] + \theta_1,
			\end{array}
		\end{align}
		then the maximum quantum value $I_{4\ {\rm{max}}}^{\rm QM}=10\sqrt {\frac{2}{3}}$ is obtained.

		\item $N$=6: The coefficient matrix $AS_6$ is
		\begin{align}\label{eqI6}
			AS_6&=\left(
			\begin{array}{c||cccccc}
				& A_1 & A_2 & A_3 & A_4   &A_5        &A_6\\
				\hline
				\hline
				B_1     & 1   & 1   & 1   & 1   & 1   & 1 \\
				B_2     & 1   & 1   & 1   & 1   & 1   & -1   \\
				B_3     & 1   & 1   & 1   & 1   & -2  & 0   \\
				B_4     & 1   & 1   & 1   & -3  & 0   & 0   \\
				B_5     & 1   & 1   & -2  & 0   & 0   & 0  \\
				B_6     & 1   & -1  & 0   & 0   & 0   & 0\\
			\end{array}
			\right),
		\end{align}
		and so the Bell expression $I_6$ is
		\begin{equation*}
			\begin{aligned}
				I_6&=\sum_{i,j}^{6}AS_6[i,j] \langle A_i B_j \rangle\\
				& \equiv \langle A_1B_1\rangle+\langle A_1B_2\rangle+\langle A_1B_3\rangle+\langle A_1B_4\rangle+\langle A_1B_5\rangle\nonumber\\
				&+\langle A_1B_6\rangle+\langle A_2B_1\rangle+\langle A_2B_2\rangle+\langle A_2B_3\rangle
				+\langle A_2B_4\rangle\nonumber\\
				&+\langle A_2B_5\rangle-\langle A_2B_6\rangle+\langle A_3B_1\rangle+\langle A_3B_2\rangle+\langle A_3B_3\rangle\nonumber\\
				&+\langle A_3B_4\rangle-2\langle A_3B_5\rangle+\langle A_4B_1\rangle+\langle A_4B_2\rangle+\langle A_4B_3\rangle\nonumber\\
				&-3\langle A_4B_4\rangle+\langle A_5B_1\rangle+\langle A_5B_2\rangle-2\langle A_5B_3\rangle+\langle A_6B_1\rangle\nonumber\\
				&-\langle A_6B_2\rangle \leq 12=\mathcal{C}_{\rm LHV}^6.
			\end{aligned}
		\end{equation*}
		If Alice and Bob choose the measurement directions with forms as (\ref{a_i1}) and (\ref{b_i1}), with
		\[Y=\frac{1}{\sqrt{\frac{N}{2}(\frac{N}{2}+1)}}=\frac{1}{{2\sqrt 3 }},\]
		and
		\begin{align}\label{b6}
			\theta_3&=- \arcsin \left[ {\frac{4}{{3\sqrt {11} }}} \right],
			\phi_1 = \theta_3 + \arccos \left[ {\frac{5}{{6\sqrt 3 }}} \right],\nonumber\\
			\theta_2& =  - \arccos \left[ {\frac{{ - 5}}{{2\sqrt {21} }}} \right] + \arccos \left[ {\frac{{\sqrt {83} }}{{2\sqrt {231} }}} \right],\nonumber\\
			\phi_2 &= \theta_2 + \arccos \left[ {\frac{7}{{6\sqrt 3 }}} \right],
			\theta_1 = \theta_2 - \phi_1 + \phi_2,\nonumber\\
			\phi_3 &= \theta_1 + \arccos \left[ {\frac{5}{{6\sqrt 3 }}} \right],
			\theta_0 = \phi_3 - \arccos \left[ { - \frac{4}{{3\sqrt {11} }}} \right],
		\end{align}
		then the maximum quantum value  $I_{6\ {\rm{max}}}^{\rm QM}=\dfrac{28}{\sqrt{3}}$ is obtained.
		
		\item $N$=8: The coefficient matrix $AS_8$ is
		\begin{align}\label{eqI8}
			AS_8&=\left(
			\begin{array}{c||cccccccc}
				& A_1 & A_2 & A_3 & A_4   &A_5        &A_6   &A_7    &A_8\\
				\hline
				\hline
				B_1     & 1   & 1   & 1   & 1   & 1   & 1 &1  &1\\
				B_2     & 1   & 1   & 1   & 1   & 1   &1   &1  & -1   \\
				B_3     & 1   & 1   & 1   & 1   &1    &1   & -2  & 0   \\
				B_4     & 1   & 1   & 1   &1  &1   & -3  & 0   & 0   \\
				B_5     & 1   & 1   &1   &1     & -4  & 0   & 0   & 0  \\
				B_6     & 1   &1    &1     & -3  & 0   & 0   & 0   & 0\\
				B_7     &1    &1    &-2    &0   &0   &0   &0   &0  \\
				B_8     &1    &-1   &0    &0    &0   &0   &0   &0  \\
			\end{array}
			\right),
		\end{align}
		and so the Bell expression $I_8$ is
		\begin{equation*}
			\begin{aligned}
				I_8&=\sum_{i,j}^{8}AS_8[i,j] \langle A_i B_j \rangle\leq 20=\mathcal{C}_{\rm LHV}^8.
			\end{aligned}
		\end{equation*}
		If Alice and Bob choose the measurement directions with forms as (\ref{a_i1}) and (\ref{b_i1}), with
		\[Y=\frac{1}{\sqrt{\frac{N}{2}(\frac{N}{2}+1)}}=\frac{1}{{2\sqrt 5 }},\]
		and
		\begin{align}\label{b8}
			\theta _5& =  - \arcsin \left[ {\frac{4}{{3\sqrt {19} }}} \right],
			\phi_1 = \theta_5 + \arccos \left[ {\frac{5}{{6\sqrt 5 }}} \right],\nonumber\\
			\theta_4 &=  - \arccos \left[ { - \frac{5}{{2\sqrt {39} }}} \right] + \arccos \left[ {\frac{{\sqrt {155} }}{{2\sqrt {741} }}} \right],\nonumber\\
			\theta_3 &=  - \arccos \left[ { - \sqrt {\frac{4}{{15}}} } \right] + \arccos \left[ {\frac{{235\sqrt {589}  + 53\sqrt {12445} }}{{7410\sqrt {12} }}} \right],\nonumber\\
			\phi_2 &= \theta_4 + \arccos \left[ {\frac{7}{{6\sqrt 5 }}} \right],
			\phi_3 = \theta_3 + \arccos \left[ {\frac{3}{{2\sqrt 5 }}} \right],\nonumber\\
			\theta_2 &= \theta_3 - \phi_2 + \phi_3,
			\phi_5 = \theta_2 - \theta_5 + \phi_2,
			\theta_1 = \theta_2 - \phi_1 + \phi_2,\nonumber\\
			\phi_4 &= \theta_1 - \theta_4 + \phi_1,
			\theta_0 = \phi_5 - \arccos \left[ {\frac{{ - 4}}{{3\sqrt {19} }}} \right],
		\end{align}
		
		then the maximum quantum value $I_{8\ {\rm{max}}}^{\rm QM}=12\sqrt 5$ is obtained.
		
	\end{itemize}

\begin{figure*}[tbph]
		\subfigure[N=4]{\includegraphics[width=45mm]{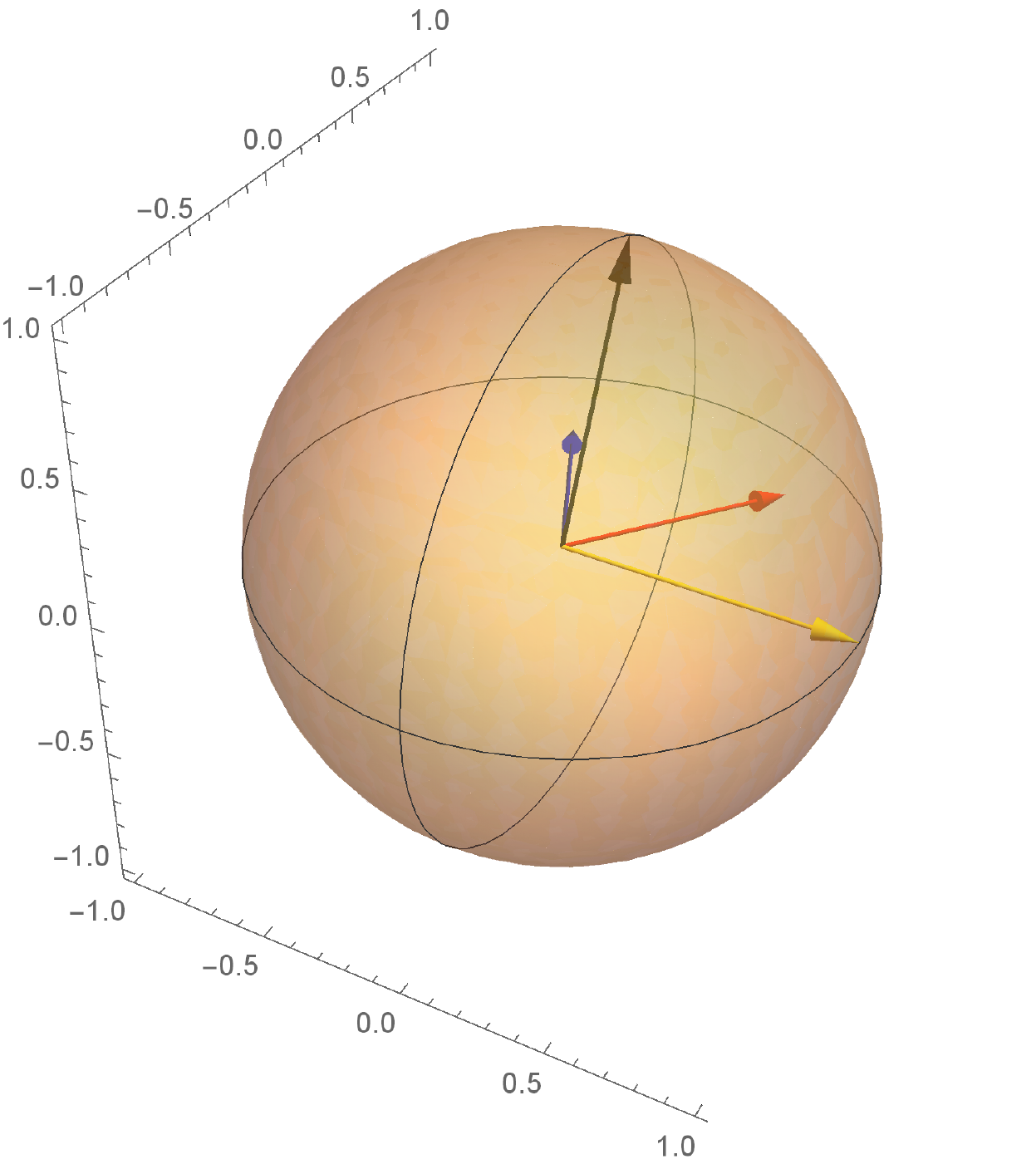}\label{fig1}}\hspace{8mm}
		\subfigure[N=6]{\includegraphics[width=45mm]{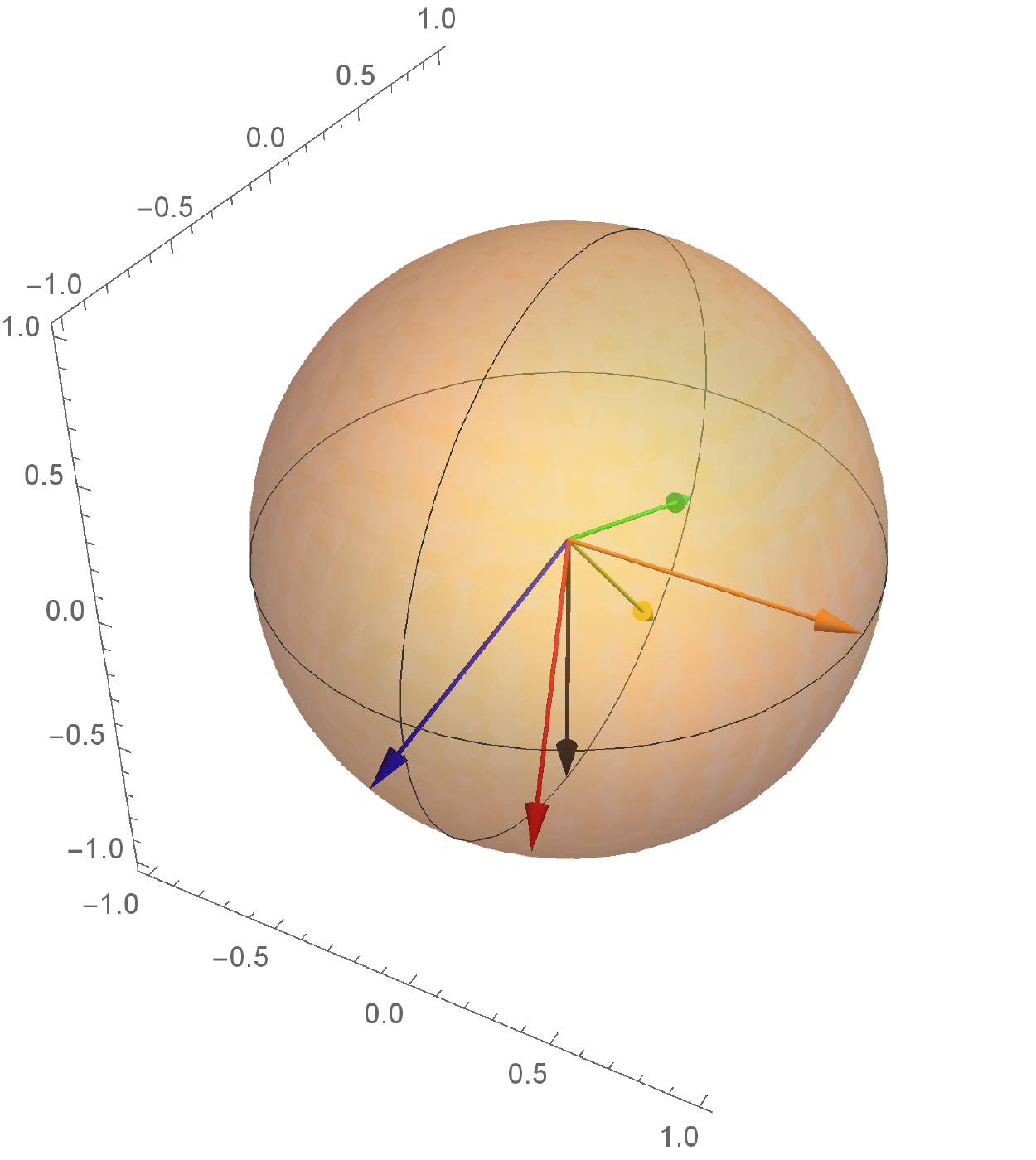}\label{fig2}}\hspace{8mm}
		\subfigure[N=8]{\includegraphics[width=45mm]{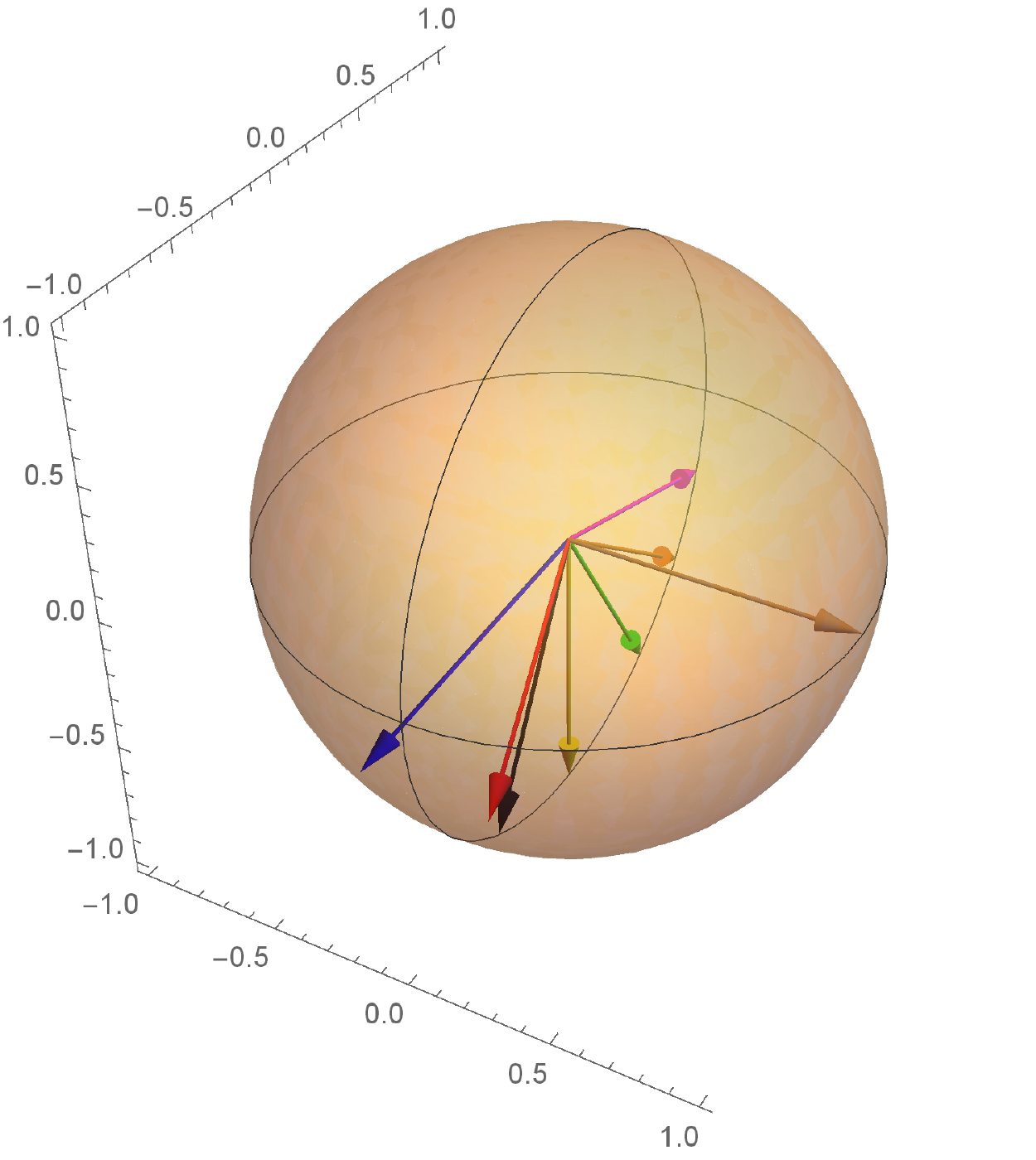}\label{fig3}}
		\caption{Bob's directions (\ref{b_i1}) with parameters $\theta_i$ being fixed as (\ref{b4}), (\ref{b6}) and (\ref{b8}) for $N=4,6,8$, respectively.}\label{figure1}
	\end{figure*}

	In Figure~\ref{figure1}, we plot  Bob's directions (\ref{b_i1}) attaining the maximum quantum value $I_{N\ {\rm{max}}}^{\rm QM}$ for $N=4,6,8$ with parameters $\theta_i$ being fixed as (\ref{b4}), (\ref{b6}) and (\ref{b8}) respectively.

	\section{Steering inequalities and comparison}\label{AS steering}
	
	In this section, we derive Einstein-Podolsky-Rosen steering inequality from the Abner Shimony inequality.
	
	In EPR-steering one considers correlations between classical variables declared by Alice but quantum expectation values found by Bob, in this sense, we call it Alice's steering Bob's particle. The steering inequality can be written as:
	\begin{align}\label{eq210-s}
		I_N^{\rm steer}\overset{{\rm LHS}}{\leq}\mathcal{C}_{\rm LHS}^N,
	\end{align}
	here
	\begin{align}\label{eq210-s1}
		I_N^{\rm steer}=\sum_{i,j}^{N}AS_N[i,j]A_i \langle B_j \rangle=\sum_{i,j}^{N}AS_N[i,j]A_i \langle \vec{\sigma}\cdot b_j \rangle,
	\end{align}
	is the EPR steering expression with $A_i\in\{1,-1\}$, and  $\mathcal{C}_{\rm LHS}^N$ is the classical bound for LHS model, which we shall determine.
	
	If Bob fixes his directions  as (\ref{b_i1}) which maximally violate the AS inequalities, then
	$\mathcal{C}_{\rm LHS}^N$ is a function of $\theta_0,\theta_1,\cdots,\theta_{N-3}$. Namely,
	\begin{align}\label{Clhs}
		&~~~~\mathcal{C}_{\rm LHS}^N(\theta_0,\theta_1,\cdots,\theta_{N-3})\nonumber\\
		&=\max\left\{I_N^{QM}:A_i\in\{1,-1\}\right\}.
	\end{align}

\begin{figure}
		\includegraphics[width=55mm]{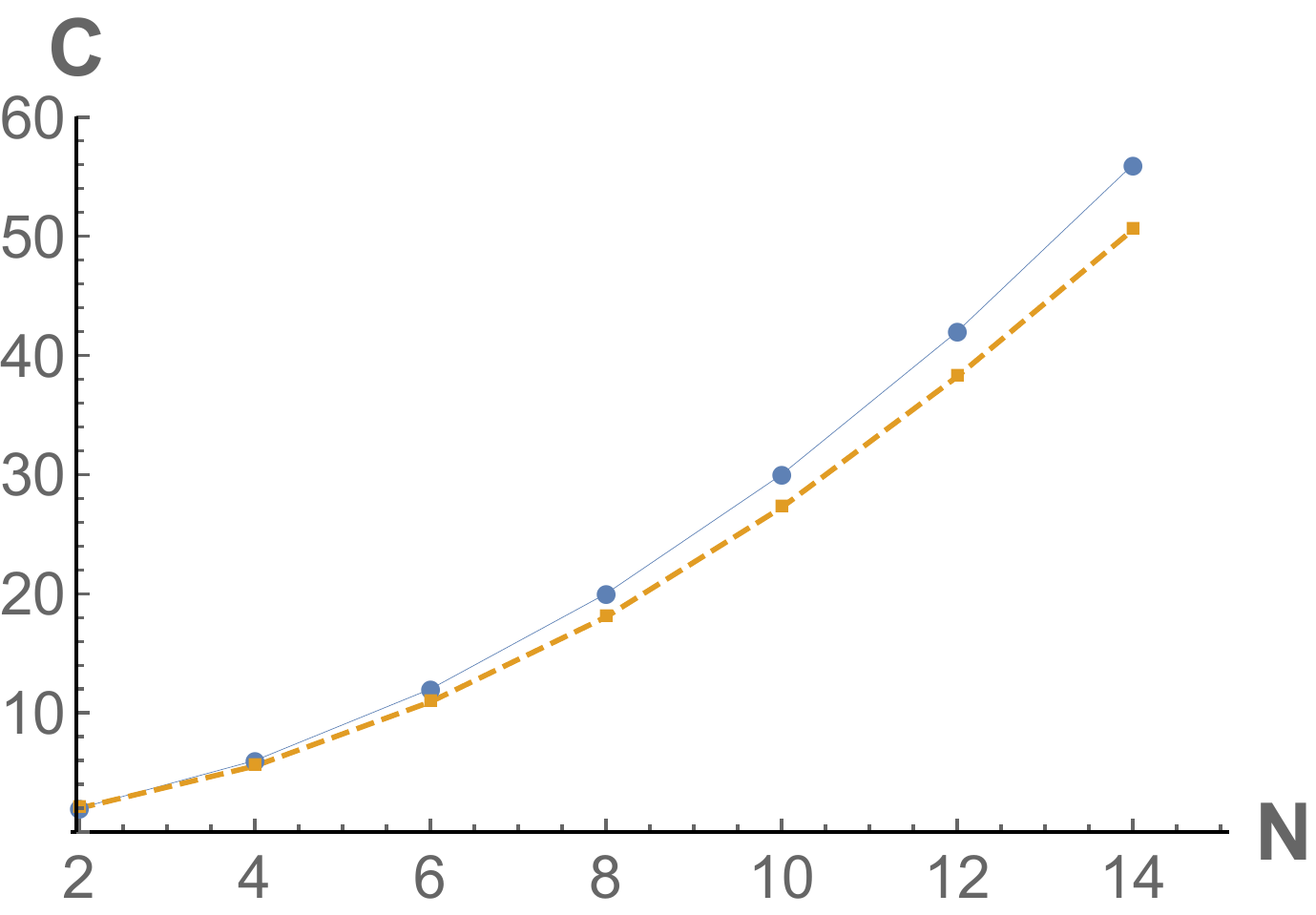}\hspace{8mm}
		\caption{\label{figure2}The $\mathcal{C}^N_{\rm LHV}$ and $\mathcal{C}^N_{\rm LHS}$ for $n=2,4,6,8,10$. The blue  dot-lines is the relationship between $\mathcal{C}^N_{\rm LHV}$ and $N$, and the  orange
			dot-lines is the relationship between $\mathcal{C}^N_{\rm LHS}$ and $N$. }
	\end{figure}

	We list as follows:
	
	\begin{itemize}
		\item $N$=2: If Bob fixes his directions as (\ref{b_41}) with $\theta_0,\theta_1$  being listed in (\ref{b2}), then
		\begin{align}
			\mathcal{C}_{\rm LHS}^2(\theta_0,\theta_1)=2=\mathcal{C}_{\rm LHV}^2.
		\end{align}

		\item $N$=4: If  Bob fixes his directions as  (\ref{b_i1}) with $\theta_0,\theta_1$ being listed in (\ref{b4}), then
		\begin{align}
			\mathcal{C}_{\rm LHS}^4(\theta_0,\theta_1)=2\sqrt{\dfrac{23}{3}}\simeq5.5377<6=\mathcal{C}_{\rm LHV}^4.
		\end{align}

		\item $N$=6: If  Bob fixes his directions as  (\ref{b_i1}) with $\theta_0,\theta_1,\theta_2,\theta_3$ being listed in (\ref{b6}), then
		\begin{align}
			\mathcal{C}_{\rm LHS}^6(\theta_0,\theta_1,\theta_2,,\theta_3)=\sqrt {\frac{{358}}{3}}\simeq10.924<12=\mathcal{C}_{\rm LHV}^6.
		\end{align}

		\item $N$=8: If  Bob fixes his directions as  (\ref{b_i1}) with $\theta_0,\cdots,\theta_5$ being listed in (\ref{b8}), then
		\begin{align}
			\mathcal{C}_{\rm LHS}^8(\theta_0,\cdots,\theta_5)=\sqrt {\frac{2(10444+\sqrt{20305})}{65}}\simeq18.0482<20=\mathcal{C}_{\rm LHV}^8.
		\end{align}

		\item $N$=10: If  Bob fixes his directions as  (\ref{b_i1}) with $\theta_0,\cdots,\theta_7$ being listed in the following: \begin{align}
			\begin{array}{l}
				\theta_0 =  - 2.5496,
				\theta_1 = 3.1742,
				\theta_2 =  - 1.9715,\\
				\theta_3 =  - 1.5541,
				\theta_4 =  - 1.0945,\\
				\theta_5 =  - 0.7886,
				\theta_6 =  - 0.5108,
				\theta_7 =  - 0.2502,
			\end{array}
		\end{align} then
		\begin{align}
			I_{10\ {\rm{max}}}^{\rm QM}=22\sqrt {\frac{{10}}{3}}
		\end{align}
		and so
		\begin{align}
			\mathcal{C}_{\rm LHS}^{10}(\theta_0,\cdots,\theta_7)\simeq27.0955<30=\mathcal{C}_{\rm LHV}^{10}.
		\end{align}
	\end{itemize}
	The comparison of $\mathcal{C}_{\rm LHV}^{N}$ and $\mathcal{C}_{\rm LHS}^{N}$ is listed in Table~\ref{table1} and plotted in Figure~\ref{figure2}.
	\begin{table}[tbp]\label{table1}
		\caption{\label{table1} $\mathcal{C}^N_{\rm LHV}$ and $\mathcal{C}^N_{\rm LHS}$ for  $N=2, 4, 6, 8, 10$. }
		\begin{ruledtabular}
			\begin{tabular}{cccccccc}
				& $N=2$ & $N=4$ & $N=6$ & $N=8$ & $N=10$ \\
				\hline
				$\mathcal{C}^N_{\rm LHV}$  & $2$ & $6$ & $12$ & $20$ & $30$\\
				$\mathcal{C}^N_{\rm LHS}$  & $2$ & $2 \sqrt{\dfrac{23}{3}}$ & $\sqrt{\frac{358}{3}}$ & $\sqrt {\frac{{2\left( {10444 + \sqrt {20305} } \right)}}{{65}}}$ & $27.0955$
			\end{tabular}
		\end{ruledtabular}
	\end{table}

	From Figure~\ref{figure2}, we can see that
	
	(i) $\mathcal{C}_{\rm LHS}^N<\mathcal{C}_{\rm LHV}^N$ for any $N>2$,
	
	(ii) for $2\leq N\leq 8$, not only both $\mathcal{C}_{\rm LHV}^N$ and $\mathcal{C}_{\rm LHS}^N$ increase with the increase of  $N$, but also the difference between $\mathcal{C}_{\rm LHV}^N$ and $\mathcal{C}_{\rm LHS}^N$  increases as $N$ does.

	If the initial quantum state of the compound system $\mathbb{C}^2\otimes\mathbb{C}^2$ is in  the Werner state
	\begin{align}\label{eq21227}
		\rho=V|\psi\rangle\langle\psi|+(1-V)\frac{I_4}{4},
	\end{align}
	where $|\psi\rangle$ denotes the singlet state (\ref{eq0110}), $I_4$ is the identity, and  $V\in[0,1]$, then
	we use $V_{\rm LHV}^N$ to denote the critical value, above which the state cannot be described by local hidden variables, and $V_{\rm LHS}^N$ to denote the critical value, above which the state cannot be described by local hidden states.
	
\begin{figure}
		\includegraphics[width=55mm]{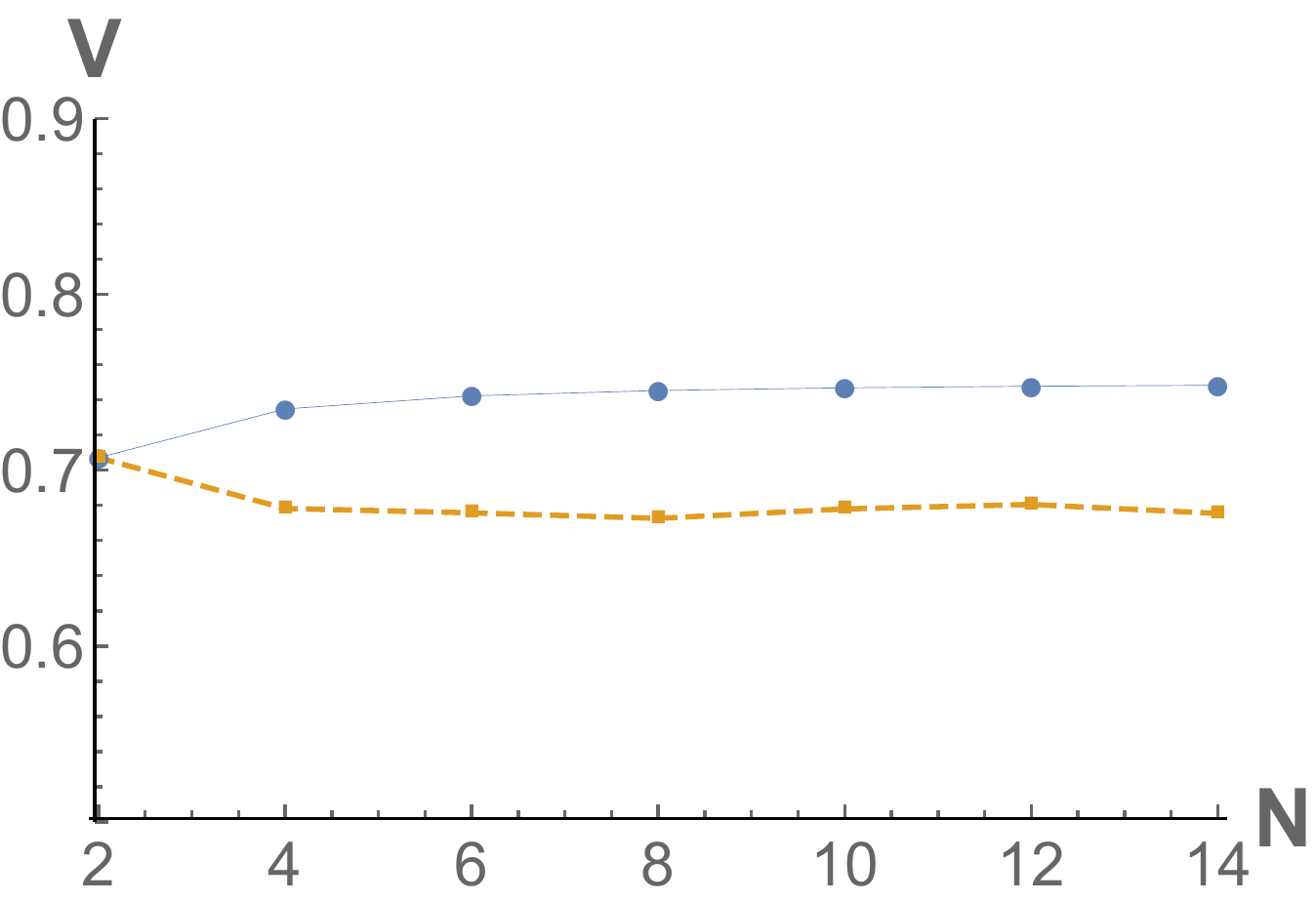}\label{fig4}\hspace{8mm}
		\caption{\label{figure3}The $V^N_{\rm LHV}$ and $V^N_{\rm LHS}$ for $N=2,4,6,8,10$. The blue dot-lines is the relationship between $V^N_{\rm LHV}$ and $N$, and the  red dot-lines is the relationship between $V^N_{\rm LHS}$ and $N$. }
	\end{figure}

	By  AS Bell inequalities (\ref{eq2140}),  AS steering inequalities (\ref{eq210-s}) and  the initial quantum state (\ref{eq21227}), we get
	\begin{align}\label{Vlhv}
		V_{LHV}^N={\mathcal{C}^N_{\rm LHV}}/{I_{N\ {\rm{max}}}^{\rm QM}},
		V_{LHS}^N={\mathcal{C}^N_{\rm LHS}}/{I_{N\ {\rm{max}}}^{\rm QM}}.
	\end{align}
	The comparison of $V_{LHV}^N$ and $V_{LHS}^N$ is listed in Table~\ref{table2} and plotted in Figure~\ref{figure3}. From Figure~\ref{figure3}, we can see that
	
	(i) $V^N_{LHS}<V^N_{LHV}$ for any $N>2$,
	
	(ii)  for $2\leq N\leq 8$, $V^N_{\rm LHV}$ increases and $V^N_{\rm LHS}$ decreases with the increase of $N$, $V_{\rm LHV}^N \rightarrow 0.75$ as $N\rightarrow\infty$, and $V_{\rm LHS}^N\leq0.7$ for any $N$.
	\begin{table}[tbp]\label{table2}
		\caption{\label{table2} We list $V^N_{\rm LHV}$ and $V^N_{\rm LHS}$ for the chained (Bell and steering) inequalities with $n=2, 4, 6, 8, 10$. Here $V^N_{\rm LHV}=\frac{{3\sqrt {N\left( {2 + N} \right)} }}{{4 + 4N}}$.}
		\begin{ruledtabular}
			\begin{tabular}{ccccccc}
				$I_N$ & $N=2$ & $N=4$ & $N=6$ & $N=8$ & $N=10$ \\
				\hline
				$V^N_{\rm LHV}$  & $\frac{1}{{\sqrt 2 }}$ & $\frac{{3\sqrt 3 }}{{5\sqrt 2 }}$ & $\frac{{3\sqrt 3 }}{7}$ & $\frac{{\sqrt 5 }}{3}$
				& $\frac{{3\sqrt {15} }}{{11\sqrt 2 }}$ \\
				$V^N_{\rm LHS}$  & $\frac{1}{{\sqrt 2 }}$ & $\frac{{\sqrt {23} }}{{5\sqrt 2 }}$ &
				$\frac{{\sqrt {179} }}{{14\sqrt 2 }}$ & $0.6726 $ & $0.6779$
			\end{tabular}
		\end{ruledtabular}
	\end{table}

	{\section{ Conclusions and discussion}}\label{discuss}
	
	In this paper, we have researched the simplest unified forms of directions attaining the maximum quantum value of the AS inequalities. Then we have derived EPR-steering inequalities from the AS inequalities, and computed their LHS bounds. Finally, by comparing the two thresholds  $V^N_{\rm LHV}$ and $V^N_{\rm LHS}$, we have shown   $V^N_{LHS}<V^N_{LHV}$ for any $N>2$. This means  that EPR-steering is a form of quantum nonlocality weaker than Bell-nonlocality, in the sense that some quantum states exist so that they violate the EPR-steering inequality but satisfy the AS-typed Bell inequality. The results are in agreement with the hierarchical structure of quantum nonlocality presented in \cite{Wiseman}. To date, we have no idea whether the directions attaining the maximum quantum value of the AS inequalities are optimal in detecting the steerability of Werner states or not.
	In the future, we shall investigate the optimization directions to detect the steerability of Werner states.

	\begin{acknowledgments}
		C.L.R. is supported by National key research and development program (No. 2017YFA0305200), the Youth Innovation Promotion Association (CAS) (No. 2015317), the National Natural Science Foundation of China (No. 11605205), the Natural Science Foundation of Chong Qing (No. cstc2015jcyjA00021), the project sponsored by SRF for ROCS-SEM (No. Y51Z030W10), the fund of CAS Key Laboratory of Quantum Information. J.L.C. is supported by National Natural Science Foundations of China (Grant No. 11475089).
	\end{acknowledgments}
	
	\vspace{8mm}

	\appendix*


\end{document}